\documentclass[a4paper,11pt]{article}
\usepackage{jcappub}
\usepackage[T1]{fontenc}
\usepackage{ae,aecompl}
\usepackage{subcaption}
\usepackage{graphicx}
\usepackage{amsmath}
\usepackage{amssymb}
\usepackage{lipsum}
\usepackage{mathtools}
\usepackage{comment}
\usepackage{multirow}
\usepackage{xcolor}
\usepackage{hyperref}
\usepackage{latexsym}
\usepackage{pifont}
\usepackage{adjustbox}
\usepackage{soul}
\usepackage{bm}
\usepackage[normalem]{ulem}
\pdfoutput=1

\usepackage{aas_macros}

\newcommand{\fwdm}{f_\mathrm{wdm}}
\newcommand{\Mwdm}{M_\mathrm{wdm}}
\newcommand{\Omm}{\Omega_\mathrm m}
\newcommand{\Omc}{\Omega_\mathrm{cdm}}
\newcommand{\Omw}{\Omega_\mathrm{wdm}}
\newcommand{\Omb}{\Omega_\mathrm b}

\newcommand{\LCDM}{$\Lambda$CDM }
\newcommand{\de}{\,\mathrm{d}}	 
\newcommand{\der}[2]{\frac{\mathrm{d}#1}{\mathrm{d}#2}} 
\def\mean#1{\left< #1 \right>} 

\title{\boldmath Mixed dark matter: matter power spectrum and halo mass function}

\author[a,b,c,d]{G.\ Parimbelli,}
\emailAdd{gabriele.parimbelli@inaf.it}
\author[b,c,d]{G.\ Scelfo,}
\emailAdd{giulio.scelfo@sissa.it}
\author[e]{S.\ K. Giri,}
\emailAdd{sambitkumar.giri@uzh.ch}
\author[e]{A.\ Schneider,}
\emailAdd{aurel.schneider@uzh.ch}
\author[f,g]{M.\ Archidiacono,}
\emailAdd{maria.archidiacono@unimi.it}
\author[h,i,j]{S.\ Camera,}
\emailAdd{stefano.camera@unito.it}
\author[b,c,d,a]{M.\ Viel}
\emailAdd{viel@sissa.it}

\affiliation[a]{INAF-OATs, Osservatorio Astronomico di Trieste,\\Via Tiepolo 11, 34131 Trieste, Italy}
\affiliation[b]{SISSA -- Scuola Internazionale Superiore di Studi Avanzati,\\Via Bonomea 265, 34136 Trieste, Italy}
\affiliation[c]{IFPU -- Institute for Fundamental Physics of the Universe,\\Via Beirut 2, 34151 Trieste, Italy}
\affiliation[d]{INFN -- Istituto Nazionale di Fisica Nucleare, Sezione di Trieste,\\Via Valerio 2, 34127 Trieste, Italy}
\affiliation[e]{Institute for Computational Science, University of Zurich,\\Winterthurerstrasse 190, 8057 Zurich, Switzerland}
\affiliation[f]{Dipartimento di Fisica, Universit\`a degli Studi di Milano,\\Via G.\ Celoria 16, 20133 Milano, Italy}
\affiliation[g]{INFN -- Istituto Nazionale di Fisica Nucleare, Sezione di Milano,\\Via G.\ Celoria 16, 20133 Milano, Italy}
\affiliation[h]{Dipartimento di Fisica, Universit\`a degli Studi di Torino,\\Via P.\ Giuria 1, 10125 Torino, Italy}
\affiliation[i]{INFN -- Istituto Nazionale di Fisica Nucleare, Sezione di Torino,\\Via P.\ Giuria 1, 10125 Torino, Italy}
\affiliation[j]{INAF -- Istituto Nazionale di Astrofisica, Osservatorio Astrofisico di Torino,\\Via Osservatorio 20, 10025 Pino Torinese, Italy}

\abstract{
We investigate and quantify the impact of mixed (cold and warm) dark matter models on large-scale structure observables.
In this scenario, dark matter comes in two phases, a cold one (CDM) and a warm one (WDM): the presence of the latter causes a suppression in the matter power spectrum which is allowed by current constraints and may be detected in present-day and upcoming surveys.
We run a large set of $N$-body simulations in order to build an efficient and accurate emulator to predict the aforementioned suppression with percent precision over a wide range of values for the WDM mass, $\Mwdm$, and its fraction with respect to the totality of dark matter, $\fwdm$.
The suppression in the matter power spectrum is found to be independent of changes in the cosmological parameters at the 2\% level for $k\lesssim 10 \ h/$Mpc and $z\leq 3.5$.
In the same ranges, by applying a {\it baryonification} procedure on both \LCDM and CWDM simulations to account for the effect of feedback, we find a similar level of agreement between the two scenarios.
We examine the impact that such suppression has on weak lensing and angular galaxy clustering power spectra.
Finally, we discuss the impact of mixed dark matter on the shape of the halo mass function and which analytical prescription yields the best agreement with simulations. We provide the reader with an application to galaxy cluster number counts.
}


\begin{document}

\maketitle
\flushbottom


\section{Introduction}
\label{sec:introduction}

In the last two decades, the $\Lambda$ cold dark matter ($\Lambda$CDM) model has settled as the standard cosmological theory.
With a small number of parameters, it provides an accurate description of the Universe at large scales.
It is well known, however, that there are considerable uncertainties and potential inconsistencies at galactic and sub-galactic scales, which could be related to the cold and collisionless nature of dark matter (DM). In particular, the \LCDM model has been claimed to overestimate the number of satellite galaxies with respect to actual observations (``missing satellites problem'', e.g.\ \cite{missing_satellites-Klypin+99, missing_satellites-Moore+99,1999MNRAS.310.1147M,2012ApJ...745..142R,2008ApJ...679.1260M,2010MNRAS.402.1995M,2018MNRAS.475.4825T}). Moreover, halo profiles are found to be much more diverse for galaxies with similar rotation curves (``diversity problem'' \cite{diversity_problem-Oman+15, diversity_problem-Tulin+18}) and with constant density cores (``cusp/core problem'', e.g.\ \cite{cusp-core,1994astro.ph..2009M,2005MNRAS.364..665D,2012MNRAS.424.1105M}) with respect to what \LCDM predicts.
Finally, it has been argued that in order to match the observed dwarf galaxy population around the Milky Way, the most massive subhalos in \LCDM simulations need to remain without stars, which seems at odds with the standard picture of galaxy formation (``too-big-to-fail problem'', e.g.\ \cite{2011MNRAS.415L..40B,2012MNRAS.420.2318L,2012ApJ...745..142R,2012AAS...21920104T,weinberg15}).

Although properly accounting for baryons can partially solve these problems (e.g.\ \cite{2011MNRAS.410.1975W,2012MNRAS.419.3304P,2012MNRAS.422.1231G}), the persistence of some issues \cite{Buckley:2017ijx}, together with the null result of laboratory searches of dark matter particles \cite{Bertone:2018krk}, motivated the study of non-cold dark matter scenarios.
Several models have been proposed in this direction (e.g.\ interacting DM \cite{diversity_problem-Tulin+18,Bohr:2020yoe}, axions \cite{Chadha-Day:2021szb}).
Among them one of the long standing solutions of the cold dark matter (CDM) crisis is provided by dark matter that decouples from the primordial plasma when still relativistic (like standard neutrinos), but soon afterwards becomes non-relativistic.
In this picture, dark matter is warm (WDM) and it is able to free-stream until late times, damping all the density perturbations below a characteristic ``free-streaming scale''.
This particular scale is in turn associated to a free-streaming mass $M_\mathrm{fs}$, a threshold below which structure formation is heavily suppressed \cite{WDM-Bode-Ostriker+01}. We can write
\begin{equation}
    M_\mathrm{fs}(\Omw,\Mwdm) = 10^{10} \left(\frac{\Omw}{0.3}\right)^{1.45} \left(\frac{h}{0.65}\right)^{3.9} \left(\frac{1 \ \mathrm{keV}}{\Mwdm}\right)^{3.45} \ M_\odot/h \,,
    \label{eq:free_streaming_mass}
\end{equation}
with  $\Mwdm$ is the WDM particle mass, $\Omw$ its density parameter and $h$ the dimensionless Hubble parameter.
Free-streaming becomes astrophysically relevant for masses of $\mathcal O$(keV) and below: typical candidates are gravitinos \cite{1984NuPhB.238..453E,1993PhLB..303..289M,2008JHEP...12..055G} and sterile neutrinos \cite{1994PhRvL..72...17D,2006PhLB..639..414S}.
Currently, the tighest constraints come from Lyman-$\alpha$ data \cite{WDM_constraint-Lyman_alpha-Irsic+17}, in particular from the combination of the XQ-100 flux power spectrum and HIRES/MIKE, which yields $\Mwdm>5.3$ keV at 95\% confidence level (C.L.).
The latter limit does however depend on a given set of astrophysical priors and on assumptions on the reionization models and on the thermal history of the intergalactic medium.
In Ref.\ \cite{strong_lensing-Gilman+19}, the authors used strong lensing to probe the free-streaming scale and the subhalo mass function: their result is $\Mwdm>5.2$ keV at 95\% C.L., while a recent study of Milky Way satellites \cite{WDM_constraint-satellites-Newton+20} found $\Mwdm>2.02$ keV at 95\% C.L., a limit that it is tightened to $\Mwdm>3.99$ keV when modelling the effect of reionization.
More recently, in Ref.\ \cite{enzi20} strong gravitational lensing with extended sources, the Lyman-$\alpha$ forest, and the number of luminous satellites in the Milky Way were analyzed jointly in a consistent framework in order to find lower limits on thermal WDM and sterile neutrinos, finding $\Mwdm>6.048$ keV at 95\% C.L..
With these numbers in mind, upcoming large-scale structure surveys, such as \textit{Euclid},\footnote{\url{https://www.euclid-ec.org/}} DESI,\footnote{\url{https://www.desi.lbl.gov/}} and LSST at the Vera C.\ Rubin Observatory,\footnote{\url{https://www.lsst.org/}} are likely not to improve much these constraints using galaxy clustering and weak lensing only, due to the fact these observables will not probe the scales where the suppression in the density perturbations and in turn on the power spectrum occurs.

However, a scenario where DM comes in two phases, a cold one and a warm one, called \textit{mixed dark matter} or \textit{cold and warm dark matter} (CWDM), constitutes an intriguing possibility and a very simple extension to the \LCDM model for which weak lensing and galaxy clustering could provide some interesting constraint.
In this picture we have, besides the WDM mass, an additional parameter, which is the WDM fraction with respect to the total DM amount:
\begin{equation}
    \fwdm = \frac{\Omw}{\Omc+\Omw},
    \label{eq:WDM_fraction}
\end{equation}
where $\Omc$ and $\Omw$ are the density parameters for CDM and WDM, respectively.
These scenarios typically display a suppression in the power spectrum, which is shallower than WDM alone, with the exception that it can occur already at relatively large scales.
For instance, the combination of a low $\Mwdm$ with a low $\fwdm$ is in agreement with observations and can in principle be detected by large-scale structure surveys.
Previous works on CWDM have focused on the physics at the halo/galactic scale, highlighting how models sharing similar free-streaming lengths but with different combinations of $\fwdm$ and $\Mwdm$ can produce halos with different properties below masses of $10^{11} \ M_\odot/h$ due to the different behaviour of the power spectra at small scales \cite{Nature_of_DM_from_MW_satellites-Anderhalden+13,Halo_structures_CWDM-Maccio+12}.
Another work \cite{boya09} used Lyman-$\alpha$ data in combination with WMAP5 \cite{wmap5} to constrain CWDM models, finding that $\fwdm<0.2$ is allowed independently of the WDM mass.
More recently, Ref. \cite{NCDM-Diamanti+17} carried out an extensive study of mixed dark matter models, fixing the WDM temperature to the Standard Model neutrino one in order to obtain the correct $\fwdm$ and also considering the fermionic or bosonic nature of these particles. 
They found, combining data from Planck, BOSS DR11 and Milky Way satellites $\fwdm<0.29 \ (0.23)$ for fermions (bosons) in the mass range $1-10$ keV and $\fwdm<0.43 \ (0.45)$ in the mass range $10-100$ keV.

This is the first of a series of two papers in which we investigate the impact of CWDM models on the main cosmological large-scale structure observables that will be probed in upcoming surveys: in particular, we will focus on the angular power spectra of galaxy clustering and cosmic shear, on the halo mass function as well as on the reconstructed (theoretical) matter power spectrum.
While for both galaxy clustering and weak lensing the link to the observable is clear, the halo mass function cannot be directly probed and further assumptions (on e.g.\ stellar content or luminosity of galaxies) residing in dark matter halos should be made.
However, using the few existing theoretical prescriptions for reproducing the mass function, we decided to provide just a qualitative investigation on observed quantities, in particular the cluster number count. 

To do so, we run a large set of high-resolution cosmological $N$-body simulations over a wide range of values in the plane $\Mwdm-\fwdm$ in order to build an emulator able to predict the suppression in the non-linear matter power spectrum with respect to $\Lambda$CDM with percent precision.
This would allow us to improve upon currently existing fitting functions for CWDM, already provided in Ref.\ \cite{Mixed_DM-Kamada+16}.
However, the focus of that paper was on strong gravitational lensing and therefore the scales involved were much smaller than this work.
Those fitting functions were obtained by using only 6 different models in the $\Mwdm-\fwdm$ plane in rather small boxes (10 Mpc/$h$) and this does not allow us to connect properly to the linear regime.
We also compare our results with the fitting formula by Ref.\ \cite{Pk_nonlinear_WDM-Viel+12} for WDM only.
In a companion paper \cite{Scelfo+in_prep} we will present a detailed Markov Chain Monte Carlo (MCMC) forecast on CWDM in future surveys.

This paper is organized as follows.
In Section \ref{sec:simulations} we present the set of simulations we run and use for our theoretical predictions.
In Section \ref{sec:power_spectrum} we focus on the matter power spectrum: we describe how the suppression due to CWDM looks like, we show how we build the emulator, we investigate possible dependencies on the cosmological parameters and check whether the baryonification effect \cite{Baryonification-Schneider+19,Baryonification-Schneider+15} is independent from the DM model assumed (\LCDM or CWDM).
In the following Sections we investigate and discuss the impact of CWDM models on some fundamental cosmological observables such as cosmic shear spectra (Section \ref{sec:angular_spectra}) and halo mass functions (Section \ref{sec:halo_mass_function}).
Finally, in Section \ref{sec:conclusions} we draw our conclusions.


\section{Simulations and dataset}
\label{sec:simulations}

\begin{figure}
\centering
\includegraphics[width=.6\textwidth]{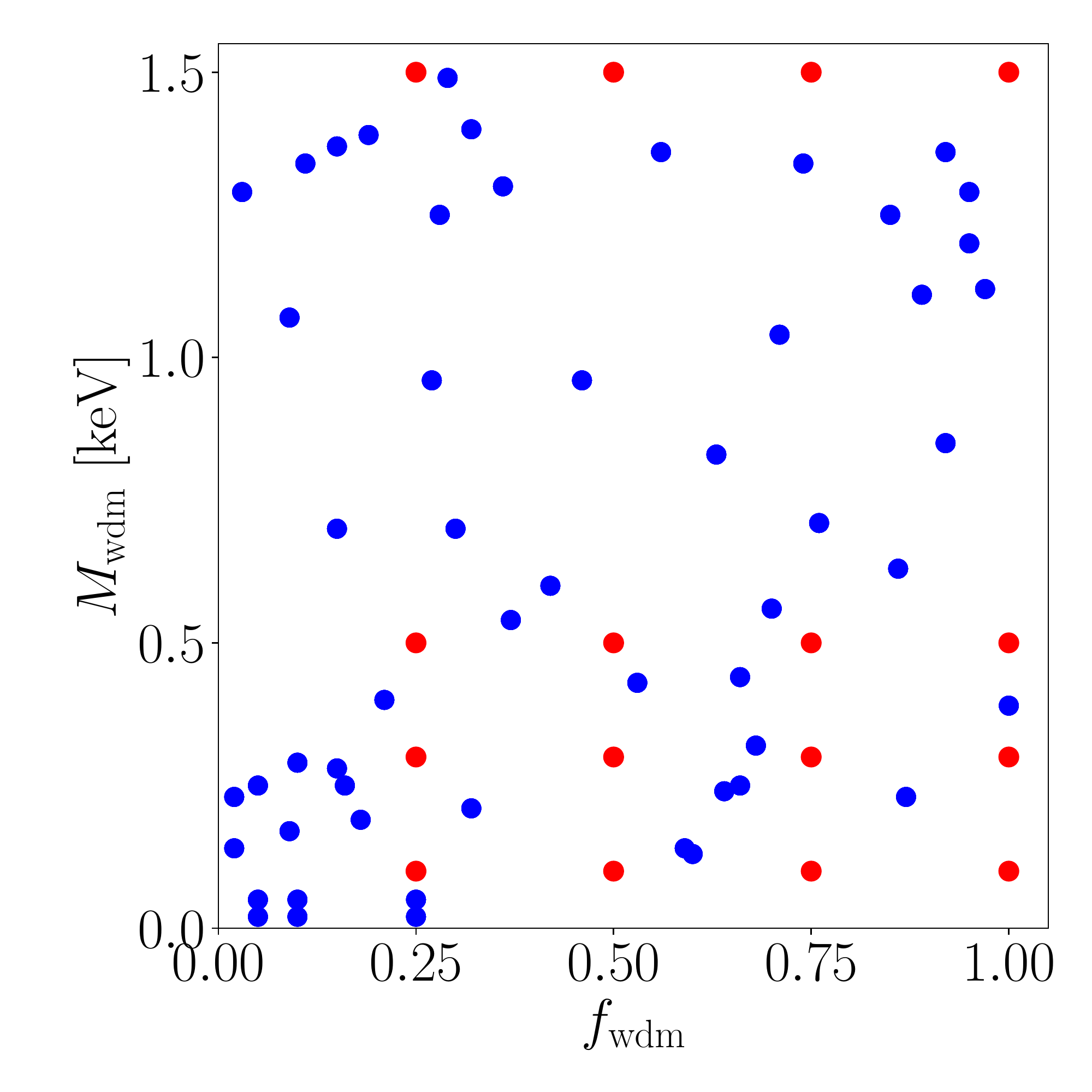}
\caption{Sampling of the $(\fwdm, \Mwdm)$ parameter space used in this work.
Red dots refer to the ``main'' set of simulations and are performed on an almost regular grid in both axes (dots relative to the $\Mwdm=3.0$ keV set are not shown here).
This set is mainly used to test the validity of theoretical predictions for the observables we consider here.
Blue dots mark the ``extra'' set, created with the purpose of populating randomly the parameter space to train the emulator for the suppression in the matter power spectrum with respect to the \LCDM model.
}
\label{fig:simulation_grid}
\end{figure}

The creation of an emulator requires a thorough sampling of the whole parameter space: this implies that a large number of different simulations has to be run for a wide range of values for $\fwdm$ and $\Mwdm$.
On the other hand, we also want to use cosmological observables for which a reliable theoretical prescription already exists, like the halo mass function \cite{HMF_WDM-Schneider+13,mass_function_WDM-Schneider+14}.
We therefore split our simulation set into two sub-sets.
The ``main'' set samples the $(\fwdm, \Mwdm)$ parameter space on an almost regular grid (20 points total).
We rely on boxes of 80 $h^{-1}$ comoving Mpc of linear size: this allows us to reconnect with the linear regime at the largest scales, while not being subjected to numerical fragmentation or resolution effects (see below) at the smallest scales of interest for future surveys ($k\lesssim 10 \ h/$Mpc).
Then, we choose the values $\fwdm = 0.25,0.5,0.75,1$ and $\Mwdm = 0.1, 0.3, 0.5, 1.5$ keV as our parameters.
The range of masses has been chosen as follows.
For masses larger than 1.5 keV the differences with \LCDM at the power spectrum level are below percent level at scales $\lesssim 10 \ h/$Mpc (we also run a set with $\Mwdm=3.0$ keV as a further check); for masses smaller than 0.1 keV and $\fwdm=1$ the suppression in the matter power spectrum starts to occur at wavenumbers comparable to the box size itself, also altering the value of $\sigma_8$ which we want to keep fixed for all of our runs.
On the other hand, the ``extra'' set randomly samples the parameter space (54 points in total) and its only purpose is to populate the training set for our matter power spectrum emulator.
Figure \ref{fig:simulation_grid} shows all the simulations that we run.
In particular, red dots correspond to the ``main'' set, while the blue ones refer to the ``extra'' set.

Our simulations have been run with the tree-particle mesh (TreePM) code \texttt{Gadget-III} \cite{GADGET-Springel+05}.
The simulations follow the gravitational evolution of $512^3$ particles from an initial redshift of $z_\mathrm{in} = 99$ to $z=0$.
All the particles were initialized assuming they were CDM and neglecting thermal velocities: 
we explicitly checked that for $\Mwdm\geq 0.3$ keV the inaccuracy introduced by this assumption does not exceed 1.5\% in the matter power spectrum overall suppression for all redshifts and scales $k\lesssim 5 \ h/$Mpc.
Thus, the differences among the different models are to be found in the initial power spectrum, computed with \texttt{CLASS} \cite{Class-Lesgourgues+11}, and therefore in the initial displacement field generated with a modified version of the \texttt{N-GenIC} software,\footnote{\url{https://github.com/franciscovillaescusa/N-GenIC_growth}} using second-order Lagrangian perturbation theory.

The fiducial cosmological parameters are set to $\Omm = \Omc + \Omw + \Omb = 0.315$, $\Omb = 0.049$, $h=0.674$, $n_\mathrm s=0.965$, $\sigma_8 = 0.811$.
With this choice of parameters, a single CDM particle has a mass of $\sim 3.3 \times 10^8 \ \ M_\odot/h$.
For each simulation we run four different realizations: two standard ones and two where the phases of the initial density field have been flipped.
We do so to reduce effects due to small-scale cross-correlations: as such, when computing matter power spectra and halo mass functions, we will always take the average of the 4 realizations.
We take 8 different snapshots, equally spaced from $z=3.5$ to $z=0$.

\begin{figure}
\centering
    \includegraphics[width=\textwidth]{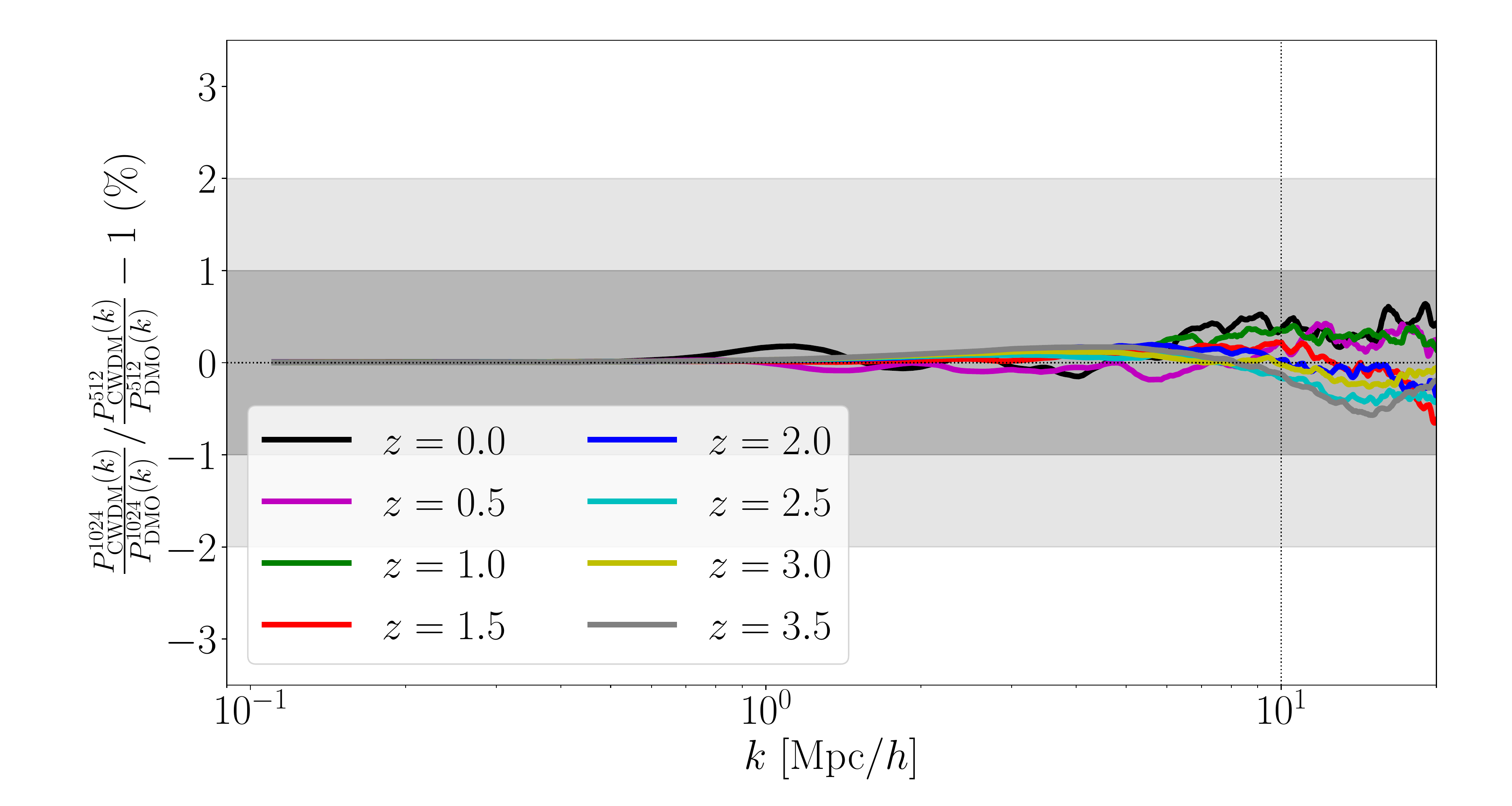}
\caption{Percentage difference in the matter power spectrum suppression for two simulations with same $\fwdm = 0.75$, $\Mwdm = 0.5$ keV and same boxsize $L=80$ Mpc/$h$ but with different number of particles (512 and 1024), to investigate the effects of resolution.
Different colors label different redshifts, the dark and light grey shaded areas represent the 1\% and 2\% regions.}
\label{fig:resolution_effects}
\end{figure}

We compute the matter power spectra using the Pylians3 code\footnote{\url{https://github.com/franciscovillaescusa/Pylians3}}.
We assign particles to a grid of size 1024 using the Cloud-In-Cell mass-assignment scheme.
This ensures that the Nyquist frequency $k_\mathrm{Nyq}\approx 40 \ h/$Mpc is much larger than the maximum $k\sim 10 \ h/$Mpc we are interested in in our analysis.
To further make sure that our results are converged at the smallest scales, we run an additional simulation set with 1024$^3$ particles while keeping the same box size.
Figure \ref{fig:resolution_effects} summarizes this test. We plot the percent difference in the suppression of the matter power spectrum when considering the simulation with 512 and 1024 particles per side.
As it can be seen, for all the redshifts considered and for all the scales of interest the difference falls within $\sim 0.5 \%$.

The halo mass function is computed with the ROCKSTAR software \cite{ROCKSTAR}.
We first identify all halos with masses larger than $10^{10} \ M_\odot/h$ through a Friends-of-Friends (FoF) algorithm, with linking length 0.28 times the mean interparticle separation.
The virial mass in this paper is defined as the mass enclosed in a region where the density is 200 times larger than the critical density of the Universe.
We operate a cut in sphericity, i.e.\ we consider only halos with axes ratios $c/a>0.24$ and $b/a>0.34$, where $a>b>c$ are the three semi-axes.
This algorithm has been shown to remove the presence of an artificial population of very elongated proto-halos which is typical in WDM scenarios but rather insignificant in \LCDM \cite{mass_function_WDM-Schneider+14,love14}.
Last, we operate a second cut in mass and keep only halos with a mass larger than 10\% of the free-streaming mass in order to avoid effects from numerical fragmentation.

\begin{figure}
\centering
    \includegraphics[width=\textwidth]{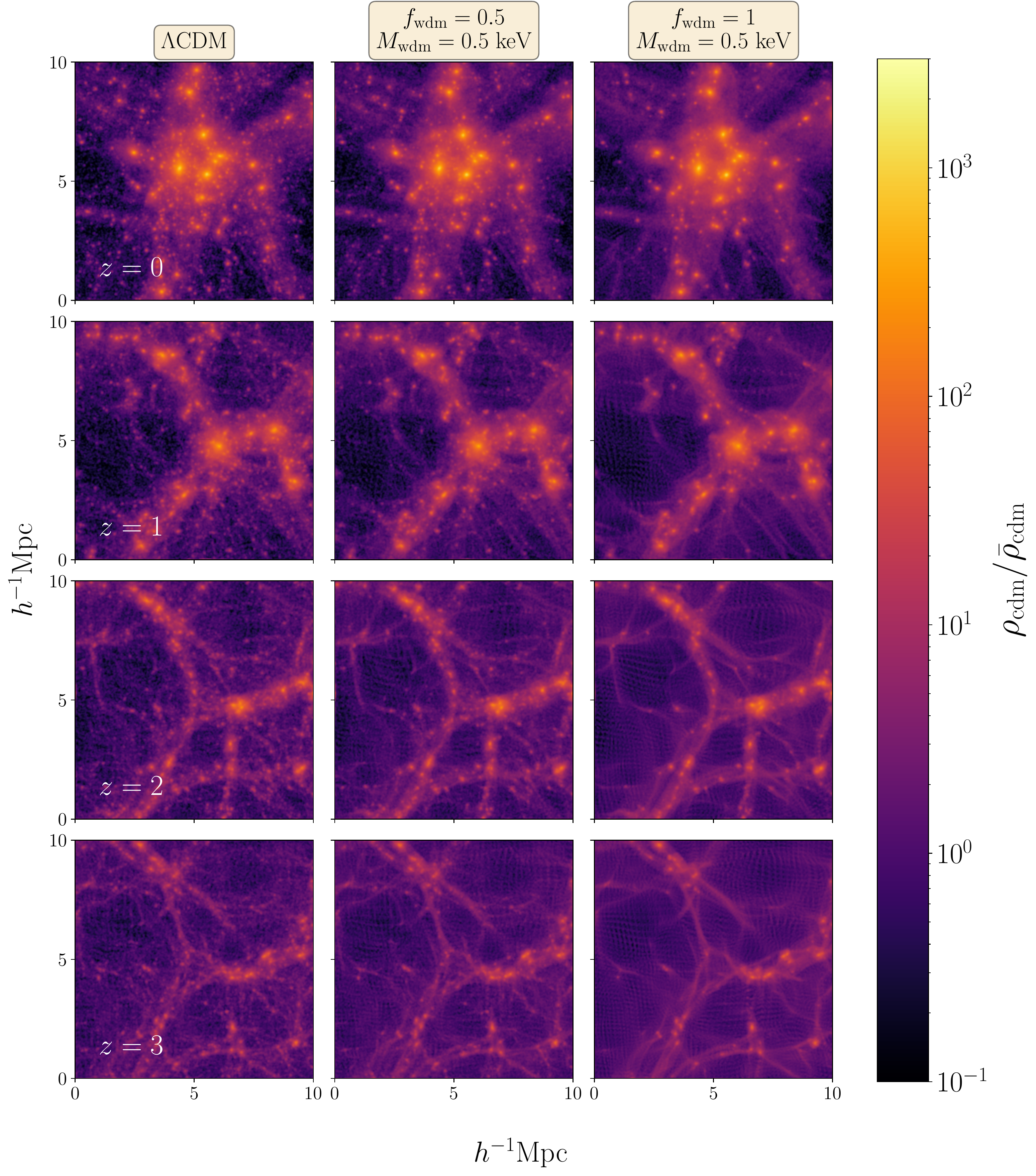}
\caption{A sample 10 Mpc/$h$ thick slice of Universe in three different models of dark matter: the left panels display the \LCDM Universe; the central panels refer to a mixed DM model with $\fwdm=0.5$ and $\Mwdm=0.5$ keV; the right panels are relative to a pure WDM model also with $\Mwdm=0.5$ keV.
From top to bottom, the four snapshots are at $z=0,1,2$ and $3$.
It is clearly visible that below the free-streaming length (roughly 1/20 of the size of the panel) structures and filaments become less prominent with increasing WDM fraction, especially at high redshift.}
\label{fig:density_plot}
\end{figure}

In Figure \ref{fig:density_plot} we show a small region ($10\times10\times10$ Mpc/$h$) of three different simulations, each with the same seed.
The left one is a pure \LCDM simulation, the one on the right is a pure WDM simulation with $\Mwdm=0.5$ keV, while the central one is a CWDM simulation also with $\Mwdm=0.5$ keV but $\fwdm = 0.5$.
Four different snapshots at $z=0,1,2,3$ are shown from top to bottom.
The free-streaming length at $z=0$ is roughly 0.4 Mpc/$h$ in pure WDM, and 0.5 Mpc/$h$ in CWDM: below this scale structures are less clustered and give rise to a suppression in the matter power spectrum and, correspondingly, on the halo mass function.

\section{Matter power spectrum}
\label{sec:power_spectrum}

We start by analyzing  the matter power spectrum.
In this Section, we first discuss how the suppression in the matter power spectrum looks like as a function of $\fwdm$, $\Mwdm$, and redshift (Section \ref{subsec:suppression}).
We compare the results of our simulations to the theoretical prediction from an emulator whose construction and testing is presented in Section \ref{subsec:emulator}.
We also show how these results extend and improve previous works in predicting the non-linear power spectrum \cite{Mixed_DM-Kamada+16,Pk_nonlinear_WDM-Viel+12,schneider12}.
In Section \ref{subsec:cosmo_dependence_suppression}, we assess possible dependencies of the suppression on the cosmological parameters.
Finally, we show how baryonic processes are independent from the DM model assumed (Section \ref{subsec:baryonification}).

\subsection{Suppression of power spectrum}
\label{subsec:suppression}

\begin{figure}
\centering
\includegraphics[width=\textwidth]{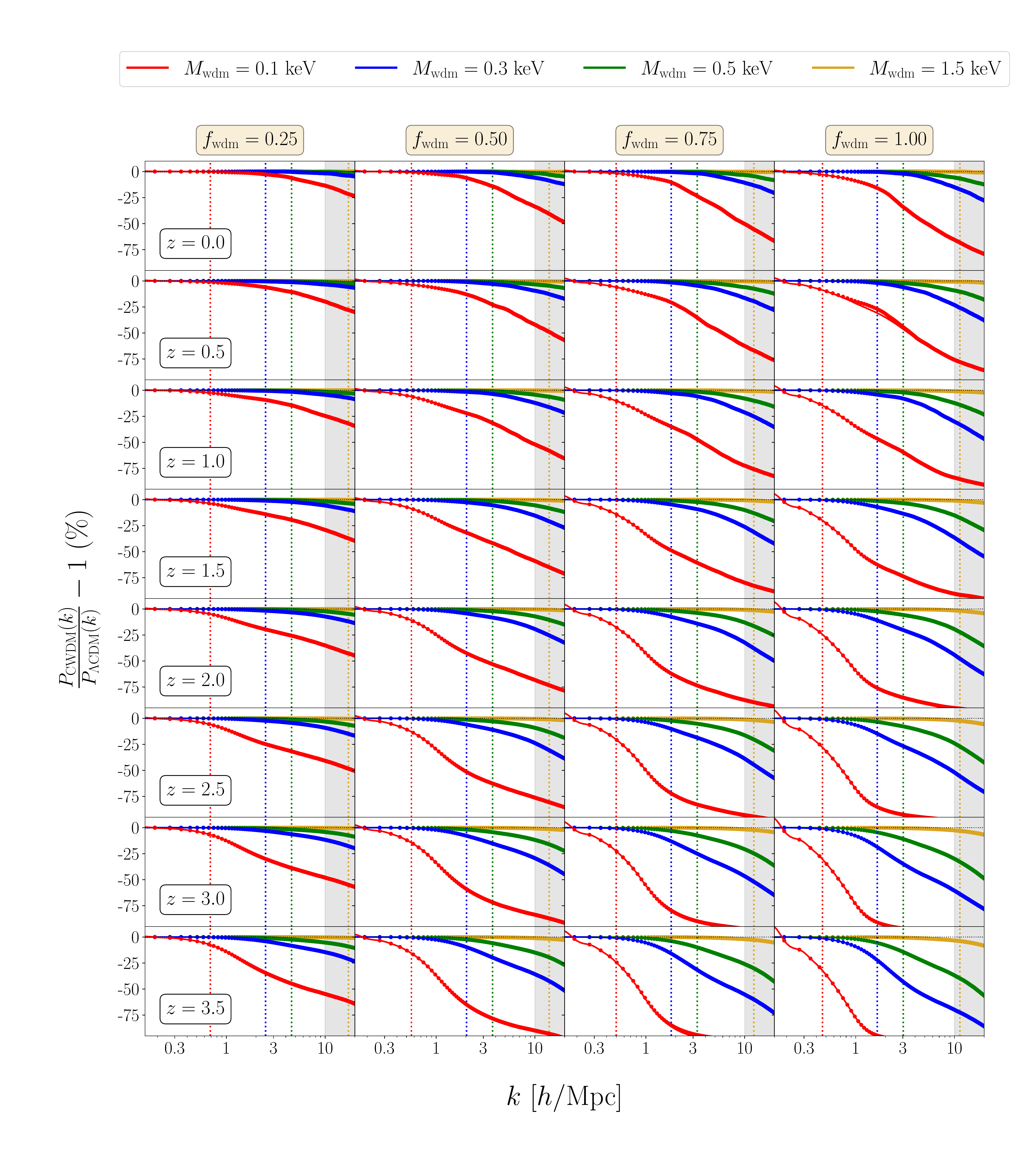}
\caption{Each subplot shows the suppression of the matter power spectrum in CWDM models with respect to \LCDM for a given redshift (rows) and a given $\fwdm$ (columns). Different colors label different $\Mwdm$: 0.1 keV in red, 0.3 keV in blue, 0.5 keV in green, and 1.5 keV in yellow.
Dots represent the measurements from the ``main'' suite of our simulations, while solid lines display the results obtained with the emulator described in Section \ref{subsec:emulator}.
Vertical colored dotted lines represent the scales at which the suppression starts to kick in, as function of $\fwdm$, $\Mwdm$, and $z$, as computed in Ref. \cite{boya09}.
The grey shaded area on the right marks the region $k>10 \ h/$Mpc, i.e. an estimate of those wavenumbers which will not be probed by upcoming large-scale structure surveys. 
}
\label{fig:Pk_suppression_CWDM}
\end{figure}

We summarize our results for what concerns matter power spectra in Figure \ref{fig:Pk_suppression_CWDM}.
Each subplot shows the suppression of power with respect to \LCDM for a given redshift (different rows) and WDM fraction (different columns).
Color-coded are the different WDM masses: red for 0.1 keV, blue for 0.3 keV, green for 0.5 keV, and yellow for 1.5 keV.
In particular, the dots represent the results from our simulations, while solid lines display the performance of our emulator (see Section \ref{subsec:emulator}).
As a reference, with the same colors each subplot shows the various scales at which the power spectrum starts to differ from the \LCDM one, computed as in Ref. \cite{boya09}.
This is referred to as \textit{free-streaming horizon}, i.e. the largest scale which is affected by WDM in cosmic history.
Finally, the grey shaded area denotes $k>10 \ h$/Mpc, a rough estimate of the scales which will not be probed by future large-scale structure surveys.
As we already mentioned, free-streaming is responsible for this suppression \footnote{We recall that we do not include thermal velocities of WDM in our simulations. However, we remark here that we explicitly checked that neglecting them has a small effect (see Section \ref{sec:simulations}), i.e. the free-streaming as imprinted in the initial power spectrum is enough to capture the suppression of the matter power spectrum.}.
The effect is more pronounced for smaller WDM masses (because of the larger thermal WDM velocities), for larger WDM fraction (as the amount of matter subject to free-streaming increases), and for increasing redshift (as the free-streaming length scales as  $1/k_\mathrm{fs}\propto(1+z)^{1/2}$ during matter domination and as the high-redshift regime is closer to linear behaviour, where primordial differences in matter power are more pronounced than in the non-linear regime).

\subsection{Emulator: building, testing and performance}
\label{subsec:emulator}

In order to find a model that best describes observations, we use a Markov chain Monte Carlo (MCMC) framework.
This method samples the parameter space, comparing theoretical predictions with data. The typical number of samples drawn in a cosmological MCMC is $\mathcal{O}(10^5)$.
However, our simulations are computationally  expensive ($>$ 5000 CPU-hours per set of parameter values).
Therefore we cannot directly explore the entire CWDM parameter space. 

In this section, we describe an emulator that can replace our simulation procedure. An emulator can be imagined as a regression model learnt from examples of our simulations, which is known as the training set.
To create the training set, we run 74 models (both the ``main'' and ``extra'' sets) by randomly sampling the $(\fwdm, \Mwdm)$ parameter space.
As described in the previous Section, we extract snapshots at 8 redshifts. The parameter space sampling is shown in Figure~\ref{fig:simulation_grid}.

Instead of directly emulating $P_\mathrm{CWDM}$, we emulate the power spectrum ratio $P_\mathrm{CWDM}/P_\mathrm{\Lambda CDM}$, so that the contribution of cosmic variance at the largest scales of the simulations is removed.
 
Our data set is very large as we have 592 power spectra, each evaluated at 886 wavenumbers. To make our emulator memory efficient, we pre-process the data set by reducing the dimensionality of our power spectra suppression data set using principal component analysis (PCA).
PCA is performed using the module provided in \texttt{scikit-learn} package \cite{scikit-learn}.   
We keep the first 20 principal components (PCs).
With these PCs, we can reconstruct our data set within an error of 2\% for $k\lesssim 10~h/\mathrm{Mpc}$.

We use the Gaussian Process Regression (GPR) \cite{GPR2006} to build our emulator. Previous works have shown that GPR is apt to emulate cosmological power spectrum, such as for non-linear matter distribution \cite{Angulo2020BACCO}, Lyman-{\ensuremath{\alpha}} forest \cite{Bird2019Lyman}, and the 21cm signal \cite{Ghara2020constraining}. We use the GPR module provided in \texttt{GPy} package \cite{gpy2014}. This module finds a model that relates the input vector $\bm{x}$ = ($\fwdm, \Mwdm, z$) to the output vector $\bm{y}$. In our work, we will emulate the coefficients of the 20 PCs.

A Gaussian process assumes any finite number of points in a parameter space to be jointly Gaussian distributed as
\begin{eqnarray}
f (\bm{x}) \sim \mathcal{N} (0, \mathcal{K}) \ ,
\end{eqnarray}
where $\mathcal{K}$ is the kernel function. This function models the similarity between the data points in the training set. There exist various choices of kernel functions \cite[see e.g.][]{GPR2006}. In our work, we use the Matern kernel \cite{minasny2005matern,GPR2006}, which contains two hyper-parameters. These hyper-parameters can be determined from the training set. 
Once we have learnt the kernel function for our parameter space, we can predict the PCs at any point in our training set. We then can reproduce the power spectra suppression from the predicted PCs. 

\begin{figure}
\centering
    \includegraphics[width=\textwidth]{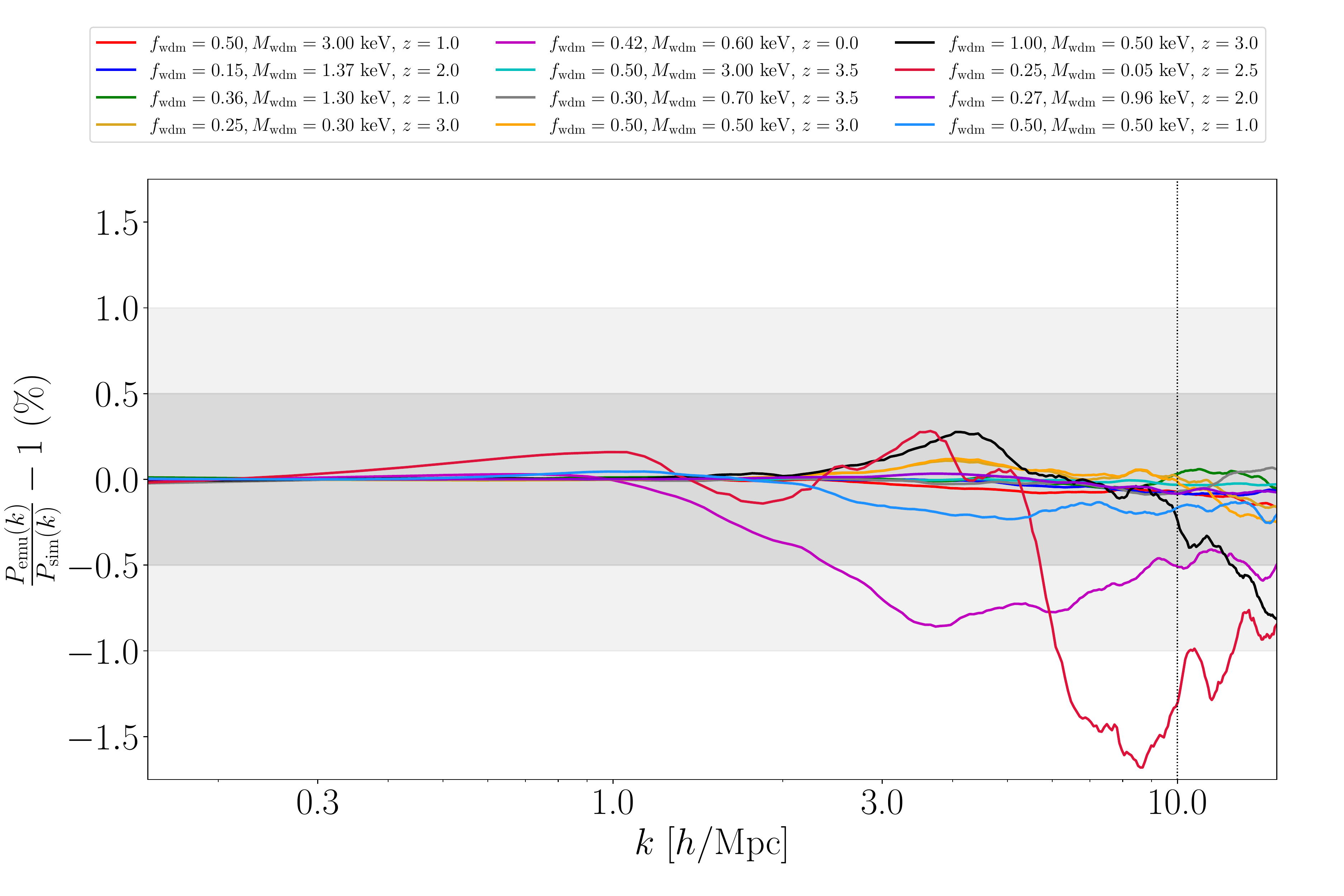}
\caption{Percentage difference between the suppression in the matter power spectrum as computed by the emulator and the one measured from randomly selected simulations of the test set. Most of the predicted suppression fall within the 1\% region; only few ($\sim$5\%) of them have a slightly worse accuracy that anyway never exceeds 1.5\%.
}
\label{fig:emulator_test}
\end{figure}

We use 90\% of our data set, which was selected randomly from the full set, to train the emulator, while we keep the remaining 10\% to test its performance.
This test set contains $\sim 60$ data points.
In Figure~\ref{fig:emulator_test}, we show the percentage difference between the emulated and simulated power spectra suppression for 12 representative data points from our test set. 
Most of the emulated power spectra suppression are within 0.5\% magnitude difference at all scales. There are quite a few power spectra suppression where the difference is larger. But this difference is only at high wavenumbers ($k\gtrsim 4 \ h$/Mpc) and does not exceed $\sim 1.5\%$.
As the emulation process has not seen the test set data points during training, a good prediction capability at these points hints that the emulator has learnt a generic model. Therefore this emulator can be used to interpolate within the parameter space where it is trained.

To have an overall picture of the performances of our emulator, we turn our attention back to the solid lines of Figure \ref{fig:Pk_suppression_CWDM}.
For all masses down to 0.1 keV we recover the correct suppression at $<2\%$ level for all $z\leq 3.5$, for all WDM fractions and masses.
Minor problems may arise at high redshift ($z\gtrsim 3$), for high fractions and small masses.
In these situations the down-turn occurs at the very same size of the box, so that some numerical errors are expected: these issues are related to the normalization of the suppression, so to the simulations themselves rather than to the emulator.
While the accuracy of the emulator with respect to the simulations remains at percent level, we detect a small ($\sim2\%$) enhancement of the power spectrum due to this effects in the range of scales comparable to the box size itself ($0.1-0.3 \ h$/Mpc).
This has a little impact on the angular power spectra of cosmic shear and galaxy clustering (see Section \ref{sec:angular_spectra}), where we have a non-physical $\sim 1\%$ enhancement at $\ell\sim 200-500$.
We remark anyway that this does not represent a problem for the purposes of this paper.
This box-size effect only interests $\fwdm\approx 1$ and the smallest masses, a region which is already by far excluded by current constraints.
Moreover, this imprecision only concerns $z\sim3$, where the typical window functions for weak lensing and angular galaxy clustering are very close to zero for benchmark future large-scale structure experiments.

Previous works have already tried to give a description of the non-linear matter power spectrum in WDM or CWDM scenario.
In particular, Refs \cite{Pk_nonlinear_WDM-Viel+12,schneider12} provided fitting formulae for the non-linear WDM suppression along the lines of its linear counterpart \cite{Pk_linear_WDM-Viel+05}.
Their claim was a 2\% agreement for $z\leq2$ and $\Mwdm \geq  0.5$ keV.
We confirm that the agreement with our simulations with $\fwdm=1$ is good, although with a slightly lower accuracy (5\%) at $z\leq2$ and $k<10 \ h/$Mpc for masses $\Mwdm\geq 0.3$ keV.
This accuracy however improves to 2\% if we limit ourselves to scales $k<3 \ h/$Mpc.
We also compared our simulations to the fitting formulae provided by Ref.\ \cite{Mixed_DM-Kamada+16}.
While these are in principle valid for both WDM and CWDM, they were obtained by comparison to $N$-body simulations with a much smaller boxsize (10 Mpc/$h$).
The fit performs similarly to the one in Ref.\ \cite{Pk_nonlinear_WDM-Viel+12} for WDM scenarios.
For CWDM we find an agreement of $\sim 5\%$ for $z\leq 2$ and $\Mwdm \geq  0.3$ keV up to scales of 10 $h$/Mpc, even though the largest deviations occur at low redshift.
Both these previous works however fail to reproduce the suppression for $\Mwdm<0.3$ keV already in the mildly non-linear regime.

\subsection{Dependence on cosmological parameters}
\label{subsec:cosmo_dependence_suppression}

We investigate possible dependencies of the CWDM suppression on cosmological parameters.
While we do not expect that the suppression depends on parameters such as $\Omb$, $h$, or $n_\mathrm s$, there might be a potential difference when we vary the overall spectrum amplitude $\sigma_8$ and the total matter content $\Omm$.
These two parameters are those that are better constrained by weak lensing and photometric galaxy clustering: in particular, the best constraints are typically achieved by combining the latter two into a single parameter $S_8=\sigma_8\left(\Omm/0.3\right)^\alpha$, where $\alpha$ is often set to 0.5 \cite{Kids1000-Heymans+20,DES-yr1-Abbott+20,DES_yr3-2021}.
Interestingly, results from the KiDS survey highlighted some tension on this parameter when analyzing the cosmic shear in Fourier space rather than in configuration space \cite{Problems_with_Kids,Kids450-power_spectrum,Kids450-correlation_function}.
Moreover, both of these values are in tension with Planck \cite{Lensing_is_low-Leauthaud+17} so that extensions to the \LCDM scenario $-$ modifications of gravity, massive neutrinos, WDM, baryon feedback (see also Section \ref{subsec:baryonification}) $-$ are typically invoked to try to solve this issue.
It is essential therefore to examine what happens in the $\Omm-\sigma_8$ plane to the suppression of the matter power spectrum also in the CWDM scenario.

We run a further set of CWDM simulations with larger and smaller $\Omm$ and $\sigma_8$ values, together with the corresponding \LCDM ones.
We choose the differences to be $\Delta\Omm = \pm 0.02$ and $\Delta\sigma_8 = \pm 0.045$.
The $\Omm$ value is $\sim 3$ times the error forecast in Ref.\ \cite{Euclid_forecast+19} by combining weak lensing and angular galaxy clustering. For $\sigma_8$, it corresponds to $\sim 12$ times this error, but we made this choice on purpose to take conservatively into account other possible effects that can alter the overall spectrum normalization, like e.g.\ massive neutrinos.
Our 8 different cosmologies therefore have $\Omm^+=0.335$,  $\Omm^-=0.295$, $\sigma_8^+=0.856$, $\sigma_8^-=0.766$ and the combinations of the two.

\begin{figure}
\centering
    \includegraphics[width=\textwidth]{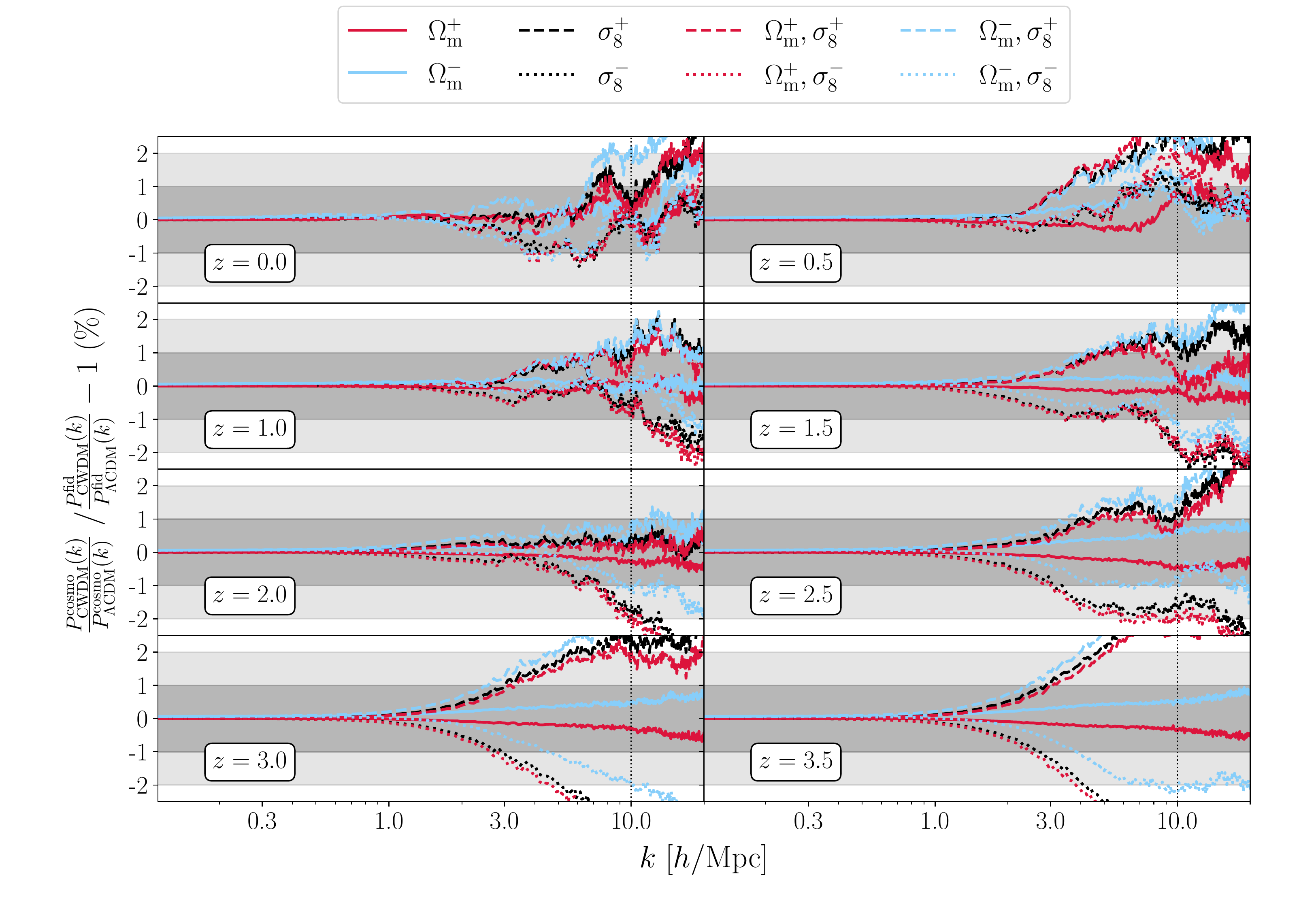}
\caption{
Dependence of the CWDM suppression in the matter power spectrum upon cosmological parameters.
We run 8 simulations with larger and smaller $\Omm$, $\sigma_8$ values and their combinations (we show here $\fwdm=0.75$, $\Mwdm=0.5$ keV for CWDM).
We compare the suppression in the CWDM power spectrum for all the different cosmologies with the same suppression in the fiducial one.
In particular, different colors of the lines are relative to different $\Omm$: red when $\Omm$ is increased to 0.335, blue when it is decreased to 0.295.
Different line styles refer to different values of $\sigma_8$: dashed for 0.866, dotted for 0.766.
The dark (light) shaded area marks the 1\% (2\%) region, the vertical dotted line marks the scale $k=10\ h/$Mpc.
}
\label{fig:cosmo_dependence}
\end{figure}

Results of this further test are plotted in Figure \ref{fig:cosmo_dependence}, where we show the ratio between the suppression of the power spectrum $P_\mathrm{CWDM}/P_\mathrm{\Lambda CDM}$ in the varied cosmologies (``cosmo'') and the one in our fiducial cosmology (``fid'').
Differences are shown in percent.
Red colors correspond to cosmologies where $\Omm$ is enhanced while light blue lines label cosmologies with a lower $\Omm$; analogously, dashed lines refer to cases with an enhanced $\sigma_8$ and dotted lines to cases where $\sigma_8$ is diminished.
As it can be noticed from the figure, the effect of varying $\Omm$ is well below 1\% even at the highest redshift we consider ($z=3.5$).
On the other hand, when we vary $\sigma_8$ the suppression becomes slightly more cosmology-dependent at high redshift, helped by the large variation we introduce in the parameter.
However, this difference is well within the 2\% level at the scales we are interested in ($k\lesssim 10 \ h/$Mpc) and the redshifts where upcoming surveys will be sensitive ($z\lesssim 2.5$).
We are therefore confident in claiming that our emulator can be used to predict the CWDM suppression also in the neighborhood of our fiducial set of cosmological parameters and, in general, in the range of interest of cosmological parameter for future surveys.

\subsection{The baryonification model in CWDM models}
\label{subsec:baryonification}

Baryon feedback has been shown to be one of the leading mechanisms capable of modifying the distribution of matter within dark matter halos up to relatively large cosmological scales (see e.g. \cite{VanDaalen+11,VanDaalen+14}).
From a cosmological point of view, it constitutes an important systematic to be taken into account \cite{Baryon_feedback_neutrinos-Parimbelli+19, Feedback_lensing-Semboloni+11, Feedback_lensing-Huang+18}, while completely ignoring its effect on the matter power spectrum can lead to a $\sim5 \sigma$ bias in the estimate of $\Omega_\mathrm m$ and $\sigma_8$ \cite{Baryonification_weak_lensing-Schneider+19, Martinelli:2020yto}.
Since the observational constraints are still poor, these phenomena are typically investigated through computationally expensive hydrodynamic simulations.
Moreover, the uncertainty caused by different AGN feedback models can reach 50\% for scales $k\leq 1 \ h/$Mpc \cite{Feedback_uncertainties-Harnois-Deraps+15}.

A novel approach circumventing the computational cost of the problem has been first proposed by Ref.\ \cite{Baryonification-Schneider+15} and subsequently improved by Ref.\ \cite{Baryonification-Schneider+19}.
In this approach, called \textit{baryonification}, baryon feedback is added on top of DM-only simulations.
In particular, the modification of the halo profiles is taken into account through the displacement of DM particles from their positions.
Such displacement depends on five parameters directly related to the physics of the gas: two parameters controlling the slope of the gas profile and its dependence on the host-halo mass ($\mu$, $\log M_c$); one parameter setting the maximum radius of gas ejection ($\theta_\mathrm{ej}$); two parameters describing the central-galactic and total stellar fractions within the halo ($\eta_\mathrm{cga}$, $\eta_\mathrm{tot}$).

\begin{figure}
\centering
    \includegraphics[width=\textwidth]{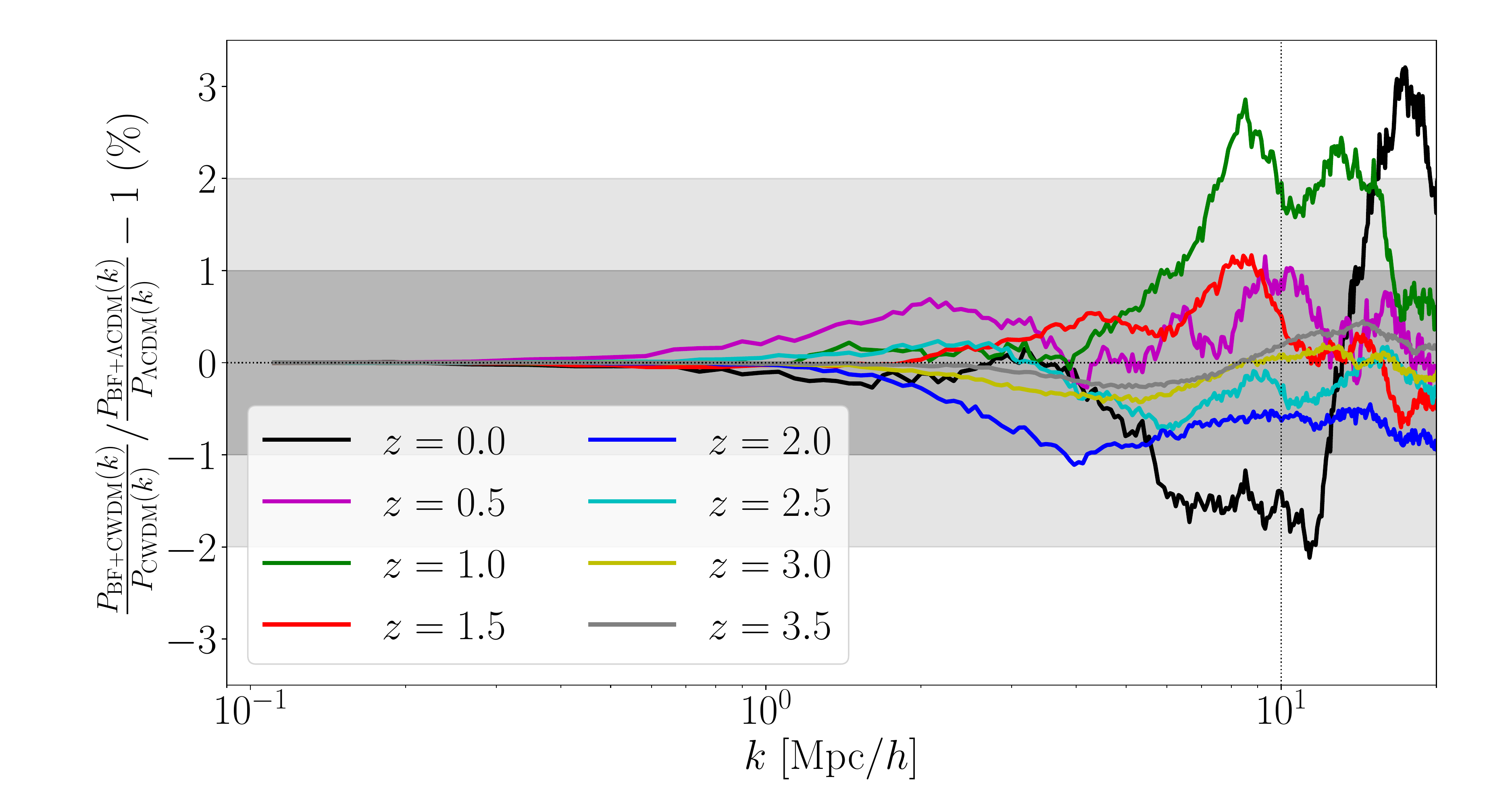}
\caption{We apply the baryonification model on top of a \LCDM as well as of a CWDM simulation and measure the suppression in the matter power spectrum due to baryon feedback in the two cases.
In this Figure we plot the ``ratio of ratios'', i.e.\ the ratio of the suppressions of $P(k)$ in the \LCDM case and in the CWDM case: in particular, $P_\mathrm{\Lambda CDM}$ and $P_\mathrm{CWDM}$ are the power spectra of the standard \LCDM and CWDM simulations, respectively, while $P_\mathrm{BF+\Lambda CDM}$ and $P_\mathrm{BF+CWDM}$ are the power spectra of the \LCDM and CWDM simulations to which we have applied the baryonification model.
Different colors refer to the different snapshots of our simulations, the grey shaded areas represent 1\% (dark) and 2\% (light) regions, respectively.
The simulation shown here has $\fwdm=0.75$ and $\Mwdm=0.5$ keV; the parameters used are $\log (M_c \ [M_\odot/h])=13.8, \mu=0.21,\theta_\mathrm{ej}=4.0, \eta_\mathrm{tot}=0.32, \eta_\mathrm{cga}=0.6$ (see main text or Ref.\ \cite{Baryonification-Schneider+19} for an insight of the baryonification parameters).
}
\label{fig:baryonification_on_cwdm}
\end{figure}

In this Section, we investigate whether the effects from baryons are separable from the suppression induced by the CWDM model. While this separability has been verified for the case of cosmologies with varying neutrino masses \cite{Mummery_2017}, it remains untested for more general CWDM scenarios. We apply the baryonification method to both our CWDM and \LCDM simulations and compare the results in terms of the relative suppression effects from baryonic feedback. The results of this analysis is illustrated in Figure \ref{fig:baryonification_on_cwdm}.
For the benchmark simulation with $\fwdm=0.75$ and $\Mwdm=0.5$ keV, we show the percent-difference of the baryonification model effect on a CWDM simulation with respect to the equivalent performed on top of the corresponding \LCDM one.
Different colors label the 8 different snapshots we took from $z=3.5$ to $z=0$.
Shaded stripes represent the 1\% (dark grey) and 2\% regions (light grey), while the scale $k=10 \ h/$Mpc is marked with a dashed black vertical line.
The parameters we used for displacing the particles ($\log (M_c \ [M_\odot/h])=13.8, \mu=0.21,\theta_\mathrm{ej}=4.0, \eta_\mathrm{tot}=0.32, \eta_\mathrm{cga}=0.6$) correspond to a model in broad agreement both X-ray observations and hydrodynamic simulations (see Ref.\ \cite{Baryonification-Schneider+19} for a more detailed discussion).

Figure \ref{fig:baryonification_on_cwdm} shows that the difference of the baryonic suppression between CDM and our benchmark CWDM scenario remains below the percent level for $k\leq 5 \ h/$Mpc (growing to 2-3 \% for $k\leq 10 \ h$/Mpc).
This is significantly smaller than the expected total baryonic suppression effect (which is of the order of $\sim10-30\%$) and of similar size of the expected precision of  $N$-body codes in the same range of scales \cite{Precision_Nbody-Schneider+16}.
We can therefore assume that CWDM suppression is independent of baryonification.
As a consequence, we can treat baryonic and CWDM power suppression effects independently, which considerably simplifies the analysis regarding the cosmological inference pipeline.

\section{Cosmic shear and galaxy clustering}
\label{sec:angular_spectra}

The next observables we focus on in the framework of CWDM models are the angular power spectra for weak lensing and galaxy clustering.
Here, we want simply to show the quantitative behaviour (together with some quantitative discussion) of the CWDM suppression on projected spectra.

In this Section we assume the Limber approximation, valid for large multipoles ($\ell\gtrsim10$) (see e.g.\ \cite{shear_full_sky-Kilbinger+17}).
In this picture, assuming a flat Universe and a single redshift bin for simplicity, the angular power spectra can be written as
\begin{equation}
C_{XY}(\ell) =
    \int \de z \, \frac{c}{H(z)} \  \frac{W_X(z) \ W_Y(z)}{\chi^2(z)} \ P\left(k=\frac{\ell+1/2}{\chi(z)},z\right),
    \label{eq:Limber}
\end{equation}
where $\chi(z)$ is the comoving distance to redsfhit $z$ and $\{X,Y\}= \{\mathrm{L,G}\}$ so that LL, GG and GL stand for cosmic shear, galaxy clustering and galaxy-galaxy lensing respectively.
The two window functions $W_\mathrm L(z)$ and $W_\mathrm G(z)$ are a measure of the lensing efficiency and the galaxy bias of the sample, respectively, and they are tightly related to the galaxy distribution in redshift $n(z)$.
The galaxy clustering window function is  given by
\begin{equation}
W_\mathrm{G}(z) = b(z) \ n(z) \ \frac{H(z)}{c},
\label{eq:galaxy_clustering_window function}
\end{equation}
where for the bias we assume the functional form by Ref.\ \cite{Flagship_bias-Tutusaus+20}, which reads
\begin{equation}
    b(z) = A+\frac{B}{1+\exp{[-(z-D)/C]}},
    \label{eq:galaxy_bias}
\end{equation}
with $\{A,B,C,D\}=\{1.0,2.5,2.8,1.6\}$.

On the other hand, the window function for cosmic shear takes into account two contributions, the former coming from the cosmological signal ($\gamma$) and the latter coming from spurious correlations of galaxy orientations coming from pairs aligned by the tidal field (IA). Hence, we have
\begin{eqnarray}
W_\mathrm{L}(z) &=& W_\gamma(z) + \mathcal F_\mathrm{IA}(z) \ W_\mathrm{IA}(z),
\label{eq:weak_lensing_window function}
\\
W_\gamma(z) &=&  \frac{3}{2} \Omm \frac{H_0^2}{c^2} \chi(z) (1+z) \int_z^\infty \de z' \ n(z') \ \frac{\chi(z')-\chi(z)}{\chi(z')},
\label{eq:shear_window function}
\\
W_\mathrm{IA}(z) &=& n(z) \ \frac{H(z)}{c},
\label{eq:IA_window function}
\end{eqnarray}

where
\begin{equation}
    \mathcal F_\mathrm{IA}(z) = -\frac{A_\mathrm{IA} \mathcal C_1 \Omega_\mathrm m}{D_1(z)} (1+z)^{\eta_\mathrm{IA}} \left[\frac{\mean{L}(z)}{L_*(z)}\right]^{\beta_\mathrm{IA}},
    \label{eq:IA_kernel}
\end{equation}
$\mathcal C_1 = 0.0134$ is fixed as it is completely degenerate with the intrinsic alignment amplitude parameter $A_\mathrm{IA}$, $D_1(z)$ is the linear growth factor.
The intrinsic alignment parameters have fiducial values $\{A_\mathrm{IA}, \eta_\mathrm{IA}, \beta_\mathrm{IA}\} = \{1.72,-0.41, 2.17\}$.
Finally, $\mean{L}(z)/L_*(z)$ is the mean luminosity of the sample in units of the typical luminosity at a given redshift. This is taken from Ref. \cite{Euclid_forecast+19}, where the authors took the luminosity functions of early and late type galaxies separately and joined them assuming a given fraction of ellipticals. The resulting function fairly reproduces fig. C.1 of Ref. \cite{Joachimi+11} in a certain $z$ range and is subsequently extrapolated to match our own redshift range \footnote{Private communication with V.F. Cardone.}.
This model is an extension of the so called non-linear alignment model \cite{Euclid_forecast+19} first introduced in Ref.\ \cite{Extended_NLA-Bridle+07}.

\begin{figure}
\centering
\includegraphics[width=\textwidth]{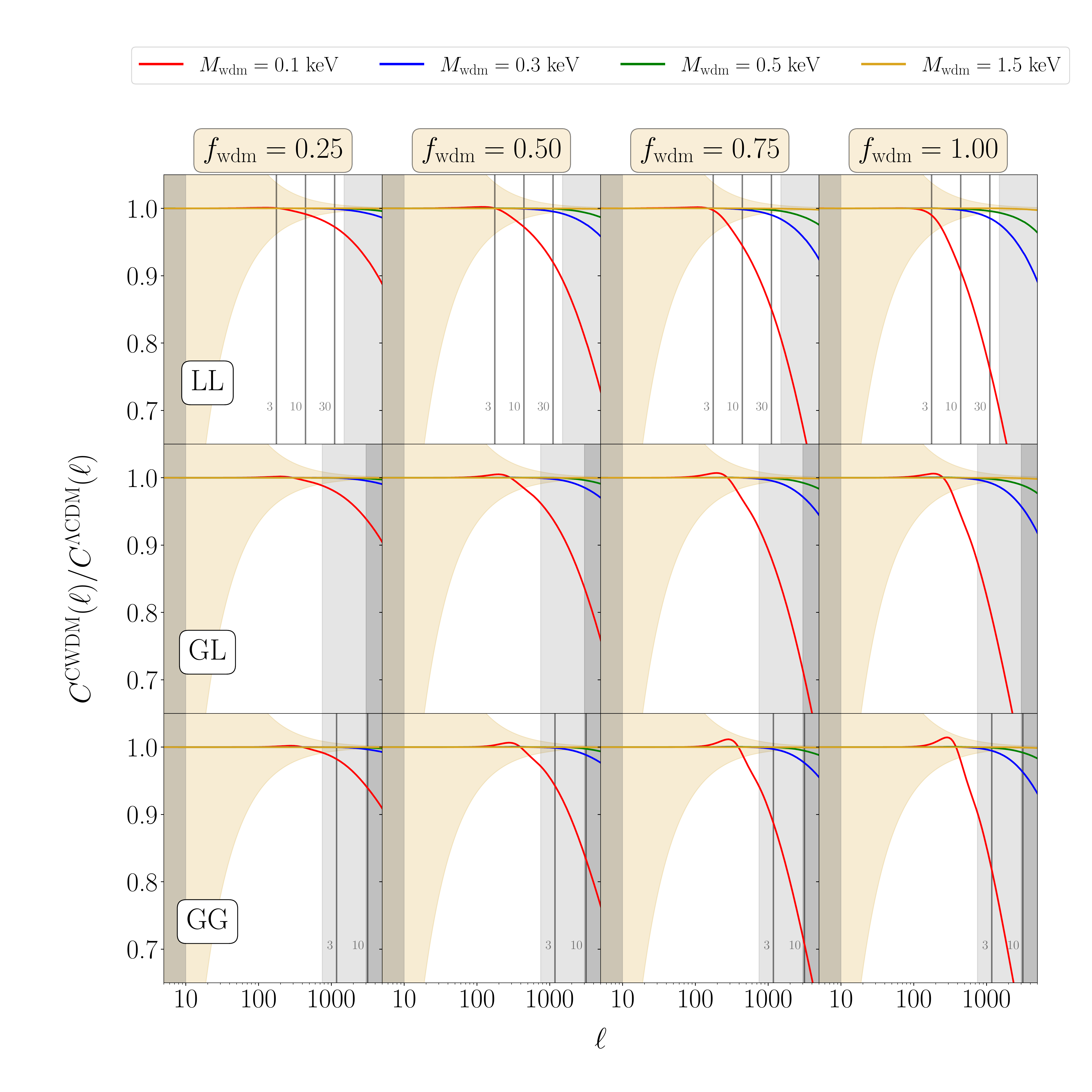}
\caption{Ratio of the angular power spectrum $C(\ell)$ in a CWDM scenario with respect to a pure \LCDM model for weak lensing (top row), galaxy-galaxy lensing (middle row), and galaxy clustering (bottom row).
Different columns report the suppression for different WDM fractions, while the WDM mass information is color-coded: red for 0.1 keV, blue for 0.3 keV, green for 0.5 keV, and yellow for 1.5 keV.
For simplicity, for this plot a single redshift bin has been used.
The golden shaded area represents cosmic variance for a survey with the same specifics of \textit{Euclid}.
The vertical lines represent the multipole at which shot/shape noise equals the cosmological signal, depending on the number of sample galaxies per square arcminute (the number written on the side of the line itself).
The vertical grey shaded areas remove the multipole regions that are likely not be used in the cosmological exploitation (see text for details): the light and dark area represent a pessimistic and an optimistic setting, respectively.
In particular, $\ell_\mathrm{min}=10$, enough to ensure the validity of the Limber's approximation, while $\ell_\mathrm{max}=1500,750,750$ for LL, GL, GG in the optimistic case and $\ell_\mathrm{max}=5000,3000,3000$ for LL, GL, GG in the optimistic case, respectively.
}
\label{fig:Cl_suppression_CWDM}
\end{figure}

In Figure \ref{fig:Cl_suppression_CWDM}, we show the ratio between the angular power spectra in the CWDM scenario with respect to $\Lambda$CDM for the three different cases of weak lensing (top row), galaxy-galaxy lensing
(middle row), and galaxy clustering (bottom row).
Different columns label different WDM fractions and different colors of the solid lines refer to different WDM masses: red for 0.1 keV, blue for 0.3 keV, green for 0.5 keV, and yellow for 1.5 keV.
The presence of the two grey shaded areas represents multipoles not considered for cosmological exploitation in the \textit{Euclid} forecasts \cite{Euclid_forecast+19}: light grey areas are eliminated when we use a pessimistic range of multipoles; dark shaded areas are excluded even in the most optimistic scenario.
In particular, the lower limit is set to $\ell_\mathrm{min}=10$ due to the fact that at lower mulitpoles the Limber's approximation is not valid.
For the maximum multipole we assume $\ell_\mathrm{max}=1500,750,750$ for LL, GL, GG in the pessimistic case and $\ell_\mathrm{max}=5000,3000,3000$ for the optimistic case.
These numbers are chosen following the lines of Ref. \cite{Euclid_forecast+19}.
The values $\ell_\mathrm{max}=5000$ and 3000 are rather optimistic for upcoming surveys, as the signal-to-noise generally saturates quickly above $\ell\gtrsim 500-1000$.
On the other hand, neglecting non-Gaussian contributions (like they do) in the data covariance matrix results in an unjustified boost in the signal-to-noise.
For cosmic shear, for instance, the signal-to-noise with a full covariance matrix (i.e. including non-Gaussian contributions) up to $\ell_\mathrm{max}=5000$ is the same to the one with a Gaussian-only covariance matrix cut at $\ell_\mathrm{cut}\sim 1500$.
The golden shaded area represents the cosmic variance limit for a survey with a sky coverage of $f_\mathrm{sky}=0.363$.
Finally, the vertical lines mark the point where shot/shape noise equals the cosmological signal, depending on how many galaxies are in the sample: we show it for 3, 10, and 30 galaxies per square arcminute, which are reasonable numbers for upcoming surveys.
We assume that galaxies follow the distribution
\begin{equation}
    n(z) \propto z^2 \exp\left[-\left(\frac{z}{z_0}\right)^{3/2}\right],
    \label{eq:galaxy_distribution}
\end{equation}
with $z_0=0.636$.
The non-linear matter power spectra for \LCDM models are computed with the HMcode2020 halofit version implemented in CAMB \cite{HMcode2020}, while to account for the presence of CWDM we use the emulator we built in Section \ref{subsec:emulator}\footnote{This is the reason why for $\Mwdm=0.1$ keV and $\fwdm>0.5$ there is a small non-physical positive bump at mutipoles $\ell\sim 300$, see Section \ref{subsec:suppression} for a more detailed explanation.}.

Figure \ref{fig:Cl_suppression_CWDM} is organized in such a way that, if the underlying cosmology is $\Lambda$CDM, each model whose line falls outside the golden shaded area at multipoles lower than the ones where shape/shot noise becomes dominant can in principle be excluded.
In general and as expected, it is easier to exclude lower values of $\Mwdm$ and high values of $\fwdm$, for which the suppression of the matter power spectrum is more pronounced.
 In the optimistic scenario and accounting for a low noise (30 arcmin$^{-2}$), cosmic shear alone could be able in principle to exclude $\Mwdm\lesssim0.3$ keV for $\fwdm>0.75$.
Galaxy clustering exhibits a less pronounced suppression but galaxy bias enhances the signal enough to allow to go to higher multipoles before being dominated by shot noise: all in all, for $\fwdm>0.75$ masses smaller than $\sim 0.5$ keV can already be excluded when sampling 10 galaxies arcmin$^{-2}$.
The suppression in galaxy-galaxy lensing, finally, has an intermediate behaviour between the previous two, but it has the advantage of being noise-free: even for the lowest WDM fraction we may be able to exclude $\Mwdm<0.3$ keV.
One can also increase the signal-to-noise by dividing galaxies into more redshift bins and combining the three observables.
Of course, this plot and this analysis have a few caveats.
First, we are completely ignoring other sources of uncertainty, like super-sample covariance \cite{SSC_lensing-Barreira+18,SSC_lensing-Krause+16} or non-Gaussian contributions that can suppress the signal-to-noise especially at high multipoles.
Moreover, here we are fixing our cosmology: we expect a worsening of the posteriors when relaxing this assumption and in particular when marginalizing over parameters like $\Omm$ and $\sigma_8$.

In Ref.\ \cite{Baryon_feedback_neutrinos-Parimbelli+19}, the authors focused on the possible degeneracies between baryon feedback and massive neutrinos in cosmic shear spectra.
In particular, they found interesting degeneracy patterns between neutrino mass and both the baryon feedback parameter $\log M_c$ and the intrinsic alignment parameter.
Since massive neutrinos could be considered WDM, we expect a similar behaviour for the case of CWDM: we leave this study to a companion paper \cite{Scelfo+in_prep} where we run MCMC forecasts for CWDM models in a \textit{Euclid}-like survey, with a proper marginalization over astrophysical, nuisance and cosmological parameters.


\section{Halo mass function}
\label{sec:halo_mass_function}

The last physical quantity that we investigate is the halo mass function.
We focus on the mass function since many observable quantities are directly linked to it, for example, the galaxy mass function, the (conditional) mass function on the number of Milky Way satellites, the number of high-redshift galaxies capable of driving reionization processes and even the strong lensing signal.
Even if the accurate modelling of the observables would require
astrophysical assumptions, the underlying dark matter mass function will always be the fundamental ingredient of any theoretical effort.

The most rigorous way of deriving it is through the excursion set of peaks \cite{Bond+91,Lacey+93,Musso+12} which extends the Press \& Schechter formalism \cite{PS+74}.
In this framework, the number of halos per unit mass per unit volume can be written as
\begin{equation}
\der{n}{\ln M} = - \frac{\bar\rho}{M} \ f(\nu) \ \der{\ln \sigma}{\ln M}.
\label{eq:halo_mass_function}
\end{equation}

In the equation above, $f(\nu)$ is a universal, cosmology-independent function of the peak height $\nu=\delta_\mathrm c/\sigma(M)$, where $\delta_\mathrm c\approx1.686$ is the linearly-extrapolated spherical overdensity for collapse and $\sigma(M)$ is the root mean square mass fluctuation
\begin{equation}
\sigma^2(M) = \int \frac{\de k \ k^2}{2\pi^2} P_\mathrm{lin}(k) \ |W(kR)|^2.
\label{eq:RMS_mass_fluctuation}
\end{equation}
The mass $M$ is related to the radius $R$ depending on the kind of window function chosen.
In the \LCDM framework, the window function $W(kR)$ that smooths the density field is typically chosen to be a top-hat in configuration space, which in Fourier space translates to
\begin{equation}
    W(x) = \frac{3}{x^3}\left[\sin x - x \cos x\right].
    \label{eq:top-hat_filter}
\end{equation}
Moreover, the universal $f(\nu)$ function is often chosen to be the Sheth-Tormen one \cite{Sheth-Tormen+99,Sheth-Tormen+01}:
\begin{equation}
    f(\nu) = A \ \sqrt{\frac{2q\nu^2}{\pi}} \left[1+(q\nu^2)^{-p}\right] \ e^{-q\nu^2/2},
    \label{eq:sheth_tormen}
\end{equation}
with $A=\left[1+2^{-p} \ \Gamma(1/2-p)/\sqrt{\pi}\right]^{-1}, \ p=0.3, \ q=0.707$.
However, when dealing with free-streaming species or, in general, with models where the power spectrum has a small-scale cut-off, the top-hat filter is not the best suitable choice as it predicts an excess of low-mass halos \cite{HMF_WDM-Schneider+13,mass_function_WDM-Schneider+14,Merger_history_WDM-Benson+12}.
A sharp-$k$ filter was invoked by Ref.\ \cite{HMF_WDM-Schneider+13} to solve this problem: our findings show that despite being able to predict fairly well the low-mass suppression, this filter suffers from problems in modelling the absolute mass function.
More recently, Ref.\ \cite{mass_function_WDM_smooth_k-Leo+18} proposed the use of a smooth-$k$ filter, namely
\begin{equation}
    W(kR) = \left[1+(kR)^\beta\right]^{-1},
    \label{eq:smooth_k_filter}
\end{equation}
where $\beta$ is a free parameter that can be fitted against $N$-body simulations.
This two new filters have the advantage of being able to alleviate the issues caused by the small-scale cut-off in the linear power spectrum.
The downside of using them is that they do not have a well-defined mass associated to the filter scale, i.e. their integral over the volume diverges.
What is typically done to restore the scaling $M\propto R^3$ is to introduce a second free-parameter $c$ to be fitted against simulations such that
\begin{equation}
    M=\frac{4}{3}\pi (cR)^3.
    \label{eq:mass_radius_relation}
\end{equation}

\begin{figure}
\centering
\includegraphics[width=.97\textwidth]{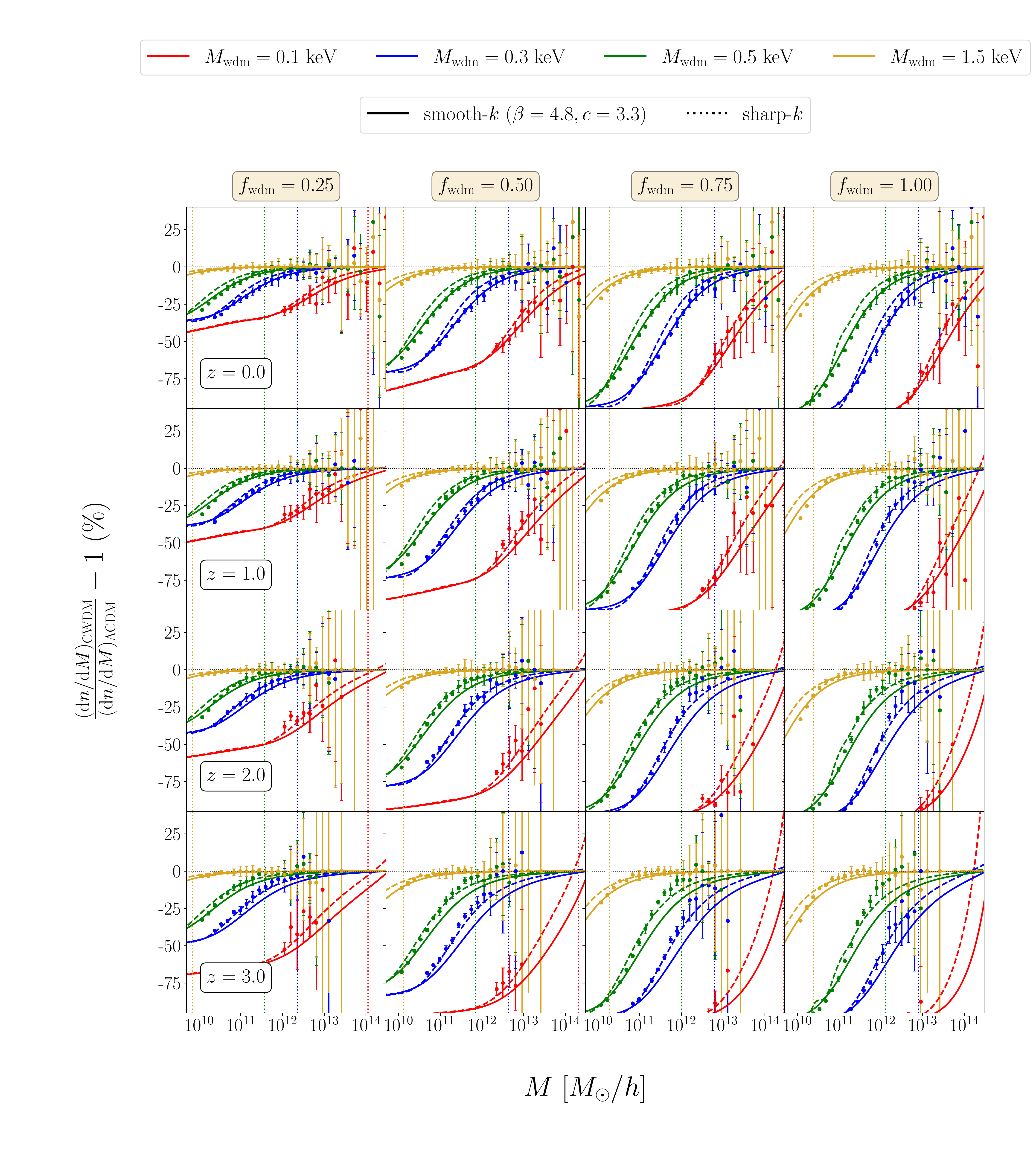}
\caption{Suppression in the halo mass function in a CWDM scenario with respect to a \LCDM cosmology, for a given redshift, WDM fraction and mass.
In particular, different columns label different WDM fractions, different rows refer to different redshifts while the WDM mass is color-coded: $\Mwdm$: 0.1 keV in red, 0.3 keV in blue, 0.5 keV in green, 1.5 keV in yellow.
Dots represent the measurements from our ``main'' suite of simulations, while error bars are taken to be the variance of the 4 realizations.
Halo mass is defined as the mass enclosed in a region where $\rho> 200 \ \rho_\mathrm{crit}$.
On the other hand, solid lines represent the theoretical prediction from the excursion set theory when a smooth-$k$ window function is used with parameters $\beta=4.8$, $c=3.3$ (see Ref.\ \cite{mass_function_WDM_smooth_k-Leo+18}); dashed lines instead represent the same but for a sharp-$k$ filter (see Ref.\ \cite{mass_function_WDM-Schneider+14}).
As a reference, we plot as vertical dotted lines the mass enclosed in a sphere of radius given by the free-streaming horizon \cite{boya09}.
}
\label{fig:HMF_suppression_CWDM}
\end{figure}

For the smooth-$k$ filter, Ref.\ \cite{mass_function_WDM_smooth_k-Leo+18} found that $\beta=4.8$ and $c=3.3$ provide a reasonable fit to $N$-body simulations, while Ref.\ \cite{dark_acoustic_oscillations-Schneider+21} obtained comparable results ($\beta=3.0$, $c=3.3$) in scenarios where dark matter produces acoustic oscillations at small scales.
We show our results in Figure \ref{fig:HMF_suppression_CWDM}.
In each subplot, we show the suppression in the halo mass function with respect to the \LCDM case.
The model used is analogous to the one described above with the only difference that we use $q=1$ in Eq. \ref{eq:sheth_tormen} \cite{HMF_WDM-Schneider+13}.
Different rows refer to different redshifts, different columns label different WDM fractions while WDM masses are color-coded: 0.1 keV in red, 0.3 keV in blue, 0.5 keV in green, and 1.5 keV in yellow.
As already mentioned in Section \ref{sec:simulations}, we define the halo mass as the mass enclosed in a radius where $\rho>200\rho_\mathrm{crit}$.
For reference, vertical dotted lines represent the mass enclosed in a sphere of radius given by the free-streaming horizon \cite{boya09}.
Solid lines represent the theoretical prediction using a smooth-$k$ filter and $(\beta,c)=(4.8,3.3)$.
We find that this combination of parameters performs slightly better than $(\beta,c)=(3.0,3.3)$, especially at low redshift.
We also show the results for a sharp-$k$ filter with $c=2.5$ (dashed lines) \cite{HMF_WDM-Schneider+13}. For the case of CWDM, we conclude that both approaches work similarly well, with the smooth-$k$ mass function providing slightly better than the sharp-$k$ mass function at low and slightly worse at higher redshifts.
Finally, it may be noticed that at small $\Mwdm$ and large $\fwdm$ the halo mass function for CWDM models assumes larger values than the \LCDM one. This once again comes from the fact that we chose to parametrize the power spectrum amplitude with $\sigma_8$ rather than for $A_\mathrm s$ and it is connected to the non-physical ``bump'' we were discussing in Sections \ref{subsec:emulator} and \ref{sec:angular_spectra}.

We want to further address the discussion on the halo mass function by linking it to actual observables, focusing on the cluster number counts.
The cluster abundance is particularly helpful in breaking the degeneracy between $\Omm$ and $\sigma_8$ thus providing tight constraints on these two parameters (see e.g.\ \cite{cluster_abundance-Wang+98,cluster_abundance-Allen+11,cluster_number_counts-Sartoris+16}).
We want now to qualitatively investigate which CWDM models will be excluded in upcoming cluster surveys.
In the simple model we consider, the cumulative number of galaxy clusters of mass larger than a threshold $M_\mathrm{th}$ is given by
\begin{equation}
N(>M_\mathrm{th}) = \int_0^{z_\mathrm{max}} \de z \ \frac{\de V}{\de z} \int_{M_\mathrm{th}}^\infty \de M \ \der{n}{M}(z),
\label{eq:cluster_number_count}    
\end{equation}

where $\de V/\de z=4\pi f_\mathrm{sky} \ c/H(z) \ \chi^2(z)$, with $f_\mathrm{sky}=0.363$, and we fix $z_\mathrm{max}=2$ and $M_\mathrm{th}=10^{13.8} \ M_\odot/h$ \cite{cluster_number_counts-Sartoris+16}.
We compute this quantity for a large set of CWDM masses, generating with \texttt{CLASS} linear matter power spectra up to very small WDM masses (even below 0.1 keV) and smoothing them with a smooth-$k$ window function with parameters $(\beta,c)=(4.8,3.3)$.
In this part of the analysis, for all the models we decide to keep $A_s$ fixed, rather than $\sigma_8$: in this way, we are able to push the WDM masses down to values for which the suppression in the power spectrum occurs at scales that influence significantly the value of $\sigma_8$ and in turn cluster abundance.

\begin{figure}[t!]
    \centering
    \includegraphics[width=\textwidth]{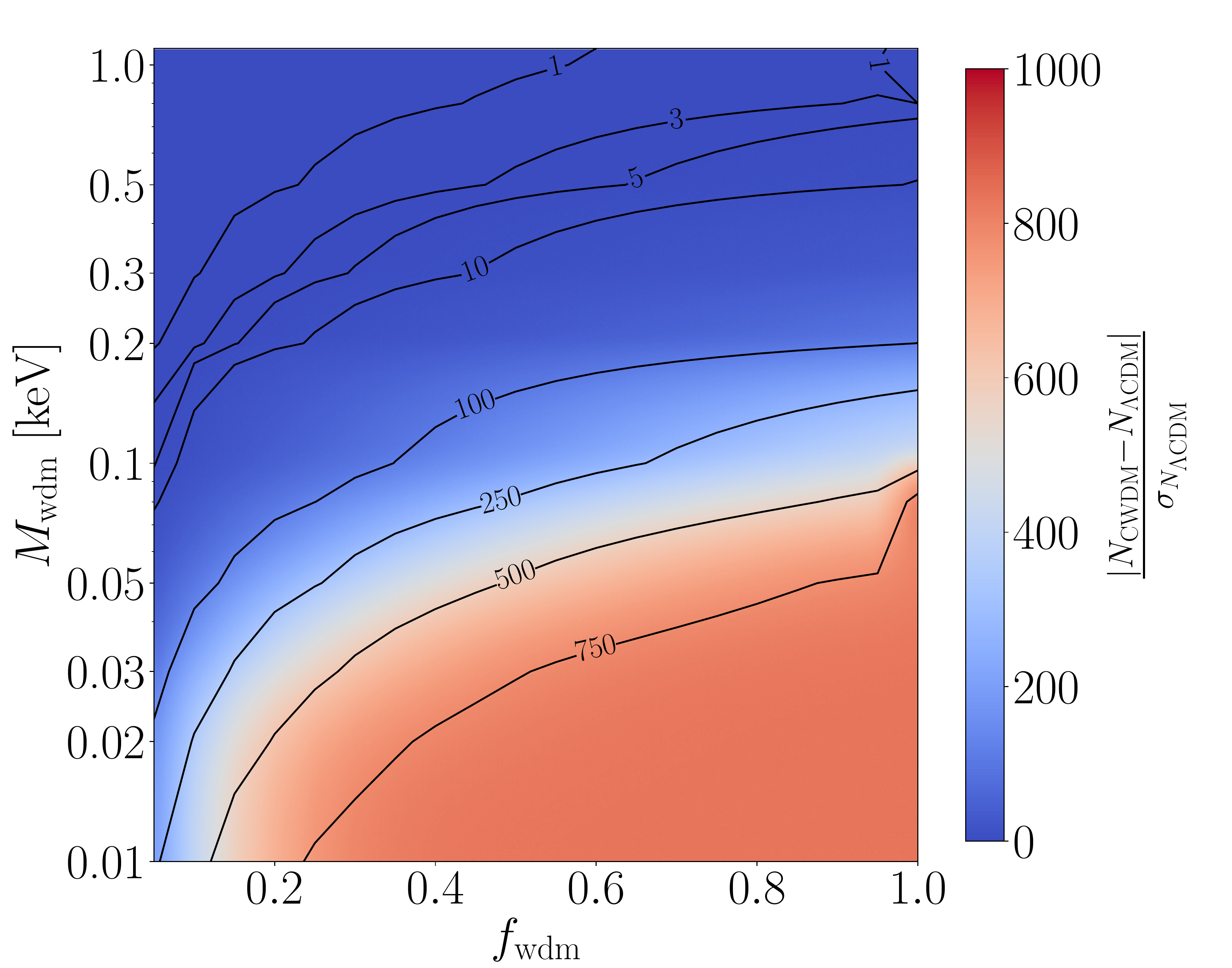}
    \caption{The Figure shows the difference between the cluster number count in CWDM models with different values of WDM fraction ($x$-axis) and WDM mass ($y$-axis) and \LCDM models, normalized by its theoretical Poisson uncertainty. The difference is color-coded; the contour lines mark the regions where such difference is 1, 3, 5, 10, 100, 250, 500, 750 times the expected error in counts assuming an underlying $\Lambda$CDM model.}
    \label{fig:cluster_MF_subfigures}
\end{figure}

In Figure \ref{fig:cluster_MF_subfigures} we show the difference between the cluster number count in CWDM scenarios and the corresponding \LCDM value, computed using eq. \ref{eq:cluster_number_count}.
Such difference is then normalized by the theoretical uncertainty on the number count itself, which is assumed to be Poissonian.
In an ideal case, models for which the difference exceeds 1-$\sigma$ will be excluded in future surveys.
However, a few caveats must be specified, as the cluster number count is subject to a number of systematics.
In particular, the complex cluster physics must be taken into account through some effective scaling relations that connect the theoretical mass function to a prediction of the distribution of clusters in the observables of the survey.
These scaling relations depend in turn on some unknown nuisance parameters (see e.g.\ \cite{cluster_number_counts-Sartoris+16,Lima+04}).
Ref.\ \cite{Salvati+20} showed how increasing the accuracy of the scaling relations leads to remarkable improvements on the constraints of the cosmological parameters; on the other hand, it causes the choice of the mass function, which for current data represents a subdominant systematic source, to become relevant.
In light of this, we keep two conservative 3-$\sigma$ and 5-$\sigma$ discrepancies as rule-of-thumb for the exclusion (or detection) of a given CWDM model.
We can crudely approximate the allowed regions as:
\begin{eqnarray}
\Mwdm &\gtrsim& \left[0.08 + 0.8 \ \fwdm^{0.82}\right] \ \mathrm{keV} \quad [3-\sigma]
\\
\Mwdm &\gtrsim& \left[0.06 + 0.7 \ \fwdm^{0.87}\right] \ \mathrm{keV} \quad [5-\sigma].
\label{eq:exclusion_CWDM_HMF}
\end{eqnarray}


\section{Discussion and conclusions}
\label{sec:conclusions}

The \LCDM model has been shown to provide an extremely accurate description of our Universe at large scales.
There are, however, remaining uncertainties at small cosmological scales, which have led to several claims of tensions between theory and observations: these  include e.g.\ the missing satellites, the too-big-to-fail, the profile diversity and the cusp-core problems.
In order to solve or alleviate these tensions warm dark matter (WDM) is often invoked.
WDM introduces a suppression in the power spectrum on scales smaller than the free-streaming length $\lambda_\mathrm{fs}$ or equivalently at masses lower than the free-streaming mass (see Eq. \ref{eq:free_streaming_mass}).
Current constraints from Milky Way satellites show that $\Mwdm>2.02$ keV \cite{WDM_constraint-satellites-Newton+20}, while Lyman-$\alpha$ studies set $\Mwdm>5.3$ keV \cite{WDM_constraint-Lyman_alpha-Irsic+17} at 95\% C.L.. With these values, the suppression in the matter power spectrum occurs at scales which are smaller than the ones probed by upcoming surveys like \textit{Euclid}.
The current constraints do not forbid  the intriguing possibility that dark matter exists in two phases, a cold one and a warm one.
This scenario, called \textit{mixed dark matter} or \textit{cold-warm dark matter} (CWDM), is the object of study of this paper.

In this work, we ran a large set of cosmological $N$-body simulations spanning a wide range of parameter values in the plane $\Mwdm-\fwdm$, where $\Mwdm$ is the WDM component mass and $\fwdm$ is the WDM fraction with respect to total DM.
We used the outputs to compute the suppression of the matter power spectrum with respect to the \LCDM case and to build an emulator: this is able to predict the suppression in power with an accuracy of $\sim1.5\%$ over the range $0<\fwdm<1, \Mwdm\gtrsim0.1$ keV, improving on previous existing fitting formulae \cite{Pk_nonlinear_WDM-Viel+12,Mixed_DM-Kamada+16}.
We also tested whether the suppression depends on cosmological parameters, in particular those which weak lensing and angular galaxy clustering are most sensitive to, i.e.\ $\Omm$ and $\sigma_8$.
We showed that such dependence is always below 2\% at the scales and redshifts of interest, even for the most extreme cases we consider, with $\sigma_8$ more than 10$\sigma$ away from our fiducial value.
We also demonstrated that the difference of the baryonic suppression between CWDM and \LCDM is much smaller than the expected total suppression effect and of the same order of magnitude of the expected precision of $N$-body codes for scales $k\lesssim10 \ h/$Mpc, thus proving that baryonic effects can be treated independently from the DM model assumed.
We used the emulated suppression to qualitatively show the impact of CWDM on weak lensing and angular galaxy clustering power spectra, focusing on which combinations of WDM masses and fractions may in principle be detected in upcoming surveys.
Finally, we studied the halo mass function.
First, we confirmed that the smooth-$k$ filter prescription proposed by Ref.\ \cite{mass_function_WDM_smooth_k-Leo+18} provides a good description both of the overall halo mass function and of its suppression in the CWDM scenario.
Then, using the same prescription, we linked the halo mass function to an actual observable, the cluster number counts, performing a semi-quantitative estimate on the CWDM models which could be probed in upcoming surveys.

In a future paper \cite{Scelfo+in_prep} we plan to perform a full Monte Carlo Markov Chain analysis on synthetic data to have more realistic forecasts on the $\Mwdm$ and $\fwdm$ parameters allowed by upcoming surveys.


\section*{Acknowledgements}
GP, GS and MV are supported by the INFN PD51 INDARK grant. SG and AS are supported by the Swiss National Science Foundation via the grant PCEFP2\_181157. GP acknowledges Alessandra Fumagalli and Vincenzo Cardone for useful discussion and Nils Sch\"oneberg for providing the code for generating the initial power spectra. Simulations were performed on the Ulysses SISSA/ICTP supercomputer.

\bibliographystyle{JHEP}
\bibliography{bibtex}



\end{document}